\documentclass[aps,twocolumn]{revtex4}

\usepackage{amssymb, amsmath, amsthm, amsfonts}
\usepackage{physics}
\usepackage{color}

\usepackage{graphicx,graphics,psfrag,calc,epsfig}

\begin{document}

\title{Effect of segregation on inequality in kinetic models of wealth exchange}

\author{L. Fernandes}
\email{Lennart.Fernandes@uantwerpen.be}
\affiliation{Theory of Quantum and Complex Systems, Universiteit Antwerpen, B-2000 Antwerpen, Belgium}

\author{J. Tempere}
\affiliation{Theory of Quantum and Complex Systems, Universiteit Antwerpen, B-2000 Antwerpen, Belgium}

\begin{abstract}
Empirical distributions of wealth and income can be reproduced using simplified agent-based models of economic interactions, analogous to microscopic collisions of gas particles. Building upon these models of freely interacting agents, we explore the effect of a segregated economic network in which interactions are restricted to those between agents of similar wealth. Agents on a 2D lattice undergo kinetic exchanges with their nearest neighbours, while continuously switching places to minimize local wealth differences. A spatial concentration of wealth leads to a steady state with increased global inequality and a magnified distinction between local and global measures of combatting poverty. Individual saving propensity proves ineffective in the segregated economy, while redistributive taxation transcends the spatial inhomogeneity and greatly reduces inequality. Adding fluctuations to the segregation dynamics, we observe a sharp phase transition to lower inequality at a critical temperature, accompanied by a sudden change in the distribution of the wealthy elite.
\end{abstract}

\maketitle

\section{Introduction}

\begin{figure}[htbp]
\includegraphics[scale=1]{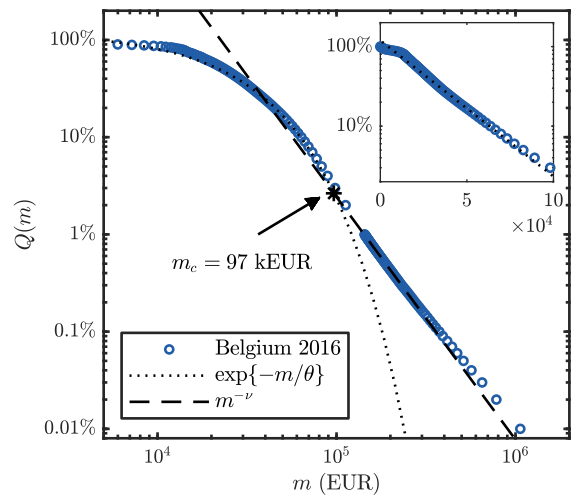}
\caption{
\label{fig:belgian_income}
(Colour online) Cumulative distribution $Q(m)=P(M>m)$ of net individual annual income $m$ in Belgium, based on tax returns of income year 2016. The data is fitted to an exponential distribution for the lowest 99\% of incomes, and to a power law for the top percentile. These regimes are illustrated by the straight lines in the log-linear inset and logarithmic main figure, respectively. The crossover point was approximated as the intersection of the fitted curves.
Data source: \cite{statbel}.
}
\end{figure}

The distribution of wealth and income is a relevant topic in any society, 
as economic inequality often lies at the heart of societal problems \cite{stiglitz_2013,piketty_2014,samuelson_2010,neckerman_2007}. Vilfredo Pareto was the first to note that income is not distributed symmetrically around a mean value, but rather follows a power law with many people at the bottom and a small wealthy elite at the top \cite{pareto_1897}. Data from a variety of countries and periods has suggested a relative insensitivity of this power law tail to economic and political details \cite{dragulescu_2001,klass_2006,abulmagd_2002}. Contemporary surveys and tax reports have led to a consensus that both wealth and income generally follow an exponential or lognormal probability distribution in the lower and middle classes, crossing over to a power law in the top percentiles \cite{dragulescu_2001,souma_2001,tao_2019,clementi_2005}:
\begin{equation}
P(m) \sim \begin{cases}
m^\gamma \exp({-m/\theta}) &\quad m<m_c,\\
m^{-(1+\nu)} &\quad m\geq m_c.
\end{cases}
\end{equation}
We illustrate this in Fig. \ref{fig:belgian_income} with new data for our home country,  Belgium. From a fit to the data, we find $\theta=25,407 \pm 12 $ EUR and $\nu=2.496 \pm 0.014$. The crossover value $m_c=97$ kEUR suggests the top 2-3\% of incomes follow a power law, while the Gini index of inequality $G\approx 0.48$ \footnote{This raw value differs from $G\approx 0.26$ reported by economic institutions such as Eurostat, who take into account additional factors such as household income \cite{eurostat}.} is in agreement with the analytical value $G=1/2$ of the exponential distribution \cite{dragulescu_2001a}.

The robust statistical properties of economic distributions and the similarity between the exponential regime and the Boltzmann distribution of energy in physical systems have led to attempts to explain these distributions with methods from statistical physics. A particularly successful path has been the development of agent-based models (ABM's), in which statistical properties are obtained through a large number of microscopic transactions in a population of economic agents, analogous to random collisions of particles in thermodynamic systems.

Besides the statistical distribution of wealth among a population discussed above, also its distribution throughout socio-economic networks is an important issue in sociological research on inequality. As social relations occur more frequently between people of similar economic status, the formation of communities of concentrated poverty and wealth can lead to a segregation of economic classes, endangering social cohesion and perpetuating existing inequalities through impaired economic mobility  \cite{neckerman_2007,mcpherson_2001,massey_1996}. Similar to the topics studied in statistical physics, this social segregation arises as a macroscopic consequence of many individuals acting on rational or irrational considerations. Such \emph{self-organizing} segregation along a categorical parameter such as ethnicity or religion was first modeled by Schelling \cite{schelling_1971}. In his model, agents on a two-dimensional lattice belong to one of two groups, and move to a different position whenever the fraction of \emph{unlike} neighbours in their local neighbourhood is higher than a critical tolerance. Surprisingly, spontaneous segregation was shown to occur even for high values of individual tolerance. The Schelling model has been studied extensively by sociologists and economists \cite{pancs_2007}, and has more recently received attention from physicists due to its rich phase diagram and similarity to physical phenomena \cite{vinkovic_2006,stauffer_solomon_2007,dallasta_2008,gauvin_2009}. Stauffer and Solomon \cite{stauffer_solomon_2007} noted the link between the Schelling model and the Ising model for ferromagnetism, and introduced thermal fluctuations in the microscopic dynamics. Socially, such fluctuations can be interpreted as the effect of external forces causing agents to move into a site which lowers their happiness.

Here, we combine the statistical models of wealth exchange and distribution on the one hand, with a dynamical model of wealth-based segregation on the other hand. In contrast to agent-based wealth exchange models studied elsewhere \cite{chatterjee_2007,yakovenko_2009}, the agents in our model interact only within their local environment, which changes continuously due to the segregation dynamics. This allows us to investigate the feedback effect of segregation, driven by wealth inequality, on the wealth inequality itself.

In Sec. \ref{sec:kinetic_models} we discuss several gas-like exchange models, containing various refinements in the modelling of economic behaviour. A mechanism for generating wealth-based segregation in a network of interacting economic agents is introduced in Sec. \ref{sec:segregation}. In Sec. \ref{sec:results} we present and discuss our results, obtained using different exchange models in conjunction with these segregation dynamics. We summarize and interpret the implications of our findings in Sec. \ref{sec:conclusion}.

\section{Kinetic models of wealth exchange}
\label{sec:kinetic_models}
We consider a population of $N$ agents, each having an initial wealth $m_i^{(t=0)}=1$. At every time step, two randomly chosen agents exchange a certain amount $\Delta m$. Their change in wealth is given by:
\begin{equation}
\begin{cases}
m_i^{(t+1)} = m_i^{(t)} + \Delta m, \\
m_j^{(t+1)} = m_j^{(t)} - \Delta m .
\end{cases}
\end{equation}
By refining the rules governing $\Delta m$, different aspects of economic behaviour can be incorporated; extensive reviews of these exchange models can be found in \cite{chatterjee_2007,yakovenko_2009}. With the aim of adding segregation to existing models, we limit ourselves to local interactions of additive exchange and focus on two relevant extensions: saving propensity and redistributive taxation. In the simplest additive model, the exchange is achieved by randomly redistributing the combined wealth of the involved agents:
\begin{equation}
\begin{cases}
m_i^{(t+1)} = \epsilon \qty( m_i^{(t)} + m_j^{(t)} ), \\
m_j^{(t+1)} = \qty(1-\epsilon) \qty( m_i^{(t)} + m_j^{(t)} ).
\end{cases}
\end{equation}
This rule intuitively corresponds to elastic collisions of gas particles. In a population where all agents can interact with each other, it leads to a steady state described by a Boltzmann law regardless of initial conditions \cite{chatterjee_2007,yakovenko_2009}. 

Multiplicative exchange processes, typically found in income from capital, have also been studied in the past. These were shown to lead to power law distributions with greater inequality and even a condensation of all wealth with one agent, implying a collapse of the economy \cite{bouchaud_2000,moukarzel_2007}.

\subsection{Saving propensity}
\label{sec:saving_model}
Realistically, people rarely spend all their money in a single exchange. We therefore introduce a saving factor $S_0$, representing the fraction of wealth that agents do not enter into an interaction \cite{chakraborti_2000}. The exchange rule then becomes
\begin{equation}
\begin{cases}
m_i^{(t+1)}=S_0 m_i^{(t)} + \epsilon \qty(1-S_0) \qty( m_i^{(t)} + m_j^{(t)} ), \\
m_j^{(t+1)}=S_0 m_j^{(t)} + \qty(1-\epsilon)\qty(1-S_0)  \qty( m_i^{(t)} + m_j^{(t)} ).
\end{cases}
\end{equation}
This individual saving behaviour prevents agents from ending up in complete poverty ($m_i=0$), altering the exponential law into a distribution peaked at finite $m$, which has been argued to be a gamma distribution \cite{patriarca_2004}. In the case of multiplicative exchange, such saving behaviour does not always prevent wealth condensation \cite{moukarzel_2007}. To reflect the fact that not everyone in society manages their money in the same way, we also study the case where some people spend a lot, while others save everything they acquire. This is achieved by assigning an individual saving factor $S_i$ to each agent, sampled uniformly from the interval $[0,1]$. Just like multiplicative exchange, this model of distributed savings leads to a power law distribution for the wealthiest percentiles of the population \cite{chatterjee_2004,patriarca_2005}, strongly resembling empirical distributions such as that in Figure \ref{fig:belgian_income}.

\subsection{Redistributive taxation}
\label{sec:tax_model}
Lastly, we include government intervention in the form of taxes. Taxation schemes vary in the object of the tax, the agents required to contribute and the way in which the tax is levied \cite{samuelson_2010}. Some taxation schemes have been studied in agent-based models \cite{dragulescu_2000,guala_2009,iglesias_2010,de_oliveira_2017}.
In the present work, we consider only simplified redistributive taxation, distinguishing between taxes on income and on wealth. In the former case, a tax is levied on each transaction by taking a fraction of the exchanged quantity $\Delta m$ and redistributing it uniformly among all agents in the population \cite{dragulescu_2000,guala_2009}. Similar to the effect of a constant saving factor, this taxation scheme suppresses the $m=0$ population, the fundamental difference being that poverty is now prevented by global intervention rather than by a change in individual behaviour. In the latter scenario, taxes are levied in proportion to an agent's total wealth at any given time. Following each exchange, we levy a certain fraction of wealth of both agents involved, and redistribute this wealth uniformly across the lattice.

\section{Wealth-based segregation}
\label{sec:segregation}
\subsection{Motivation}
The difficulty in reducing complex social dynamics to tractable systems has led to many specific models of varying complexity, each with their own assumptions and parameters. In this section we propose a new and relatively simple model of wealth-based segregation in an evolving economy, which displays rich dynamics while at the same time overcoming several limitations of existing models. 

Firstly, none of the models found in the literature consider exchanges of wealth in conjunction with segregation dynamics, instead keeping the agents' wealth fixed throughout the simulation \cite{sahasranaman_2016,benard_2007}. However, the inclusion of economic transactions is crucial to study the effect of segregation on global inequality. A second challenge arises in the incorporation of a continuous wealth variable. In the past, a common approach to both measuring and modelling economic segregation has been the division of a population into discrete \emph{classes}, allowing the use of categorical measures \cite{benard_2007,pangallo_2019,louf_2016}. While parameter-free methods have been developed to define classes in empirical data \cite{louf_2016}, any categorization in a computational model will inevitably introduce a presumptive and possibly artificial class structure. In a recent alternative approach, the Schelling model was extended to continuous wealth by treating any richer neighbours as \emph{others} \cite{sahasranaman_2016}. With the aim of modelling dynamics driven by the affordability of residences, agents in this model are drawn towards poorer neighbours. For our purpose of achieving a homophily towards agents of comparable wealth status, this model neglects the unhappiness caused by poorer neighbours, and disregards that a greater wealth difference should lead to greater unhappiness. 

To overcome these limitations, we combine both wealth exchange and segregation dynamics, driven by local differences in wealth, regardless whether neighbours are richer or poorer. In retaining the continuous character of wealth throughout our treatment, we avoid any artifacts that may arise from classification. The price to pay is that we cannot strictly speak of segregation between any distinct classes. What we mean by \emph{segregation} in the remainder of this paper is therefore no more than a macroscopic gradient of wealth, as a consequence of microscopic homophily. Nevertheless, interpreting segregation as \emph{any pattern that deviates significantly from a random distribution} \cite{louf_2016}, we may still lay claim to the terminology.

\subsection{Model definition}
Following the tradition of magnetic systems, we define a Hamiltonian,
\begin{equation}
H=\sum_{<ij>}\left( \frac{m_i-m_j}{\expval{m}}\right)^2,
\label{eq:hamiltonian}
\end{equation}
which runs over all nearest-neighbour links in the two-dimensional lattice.
The energy of the system is minimized when all local differences in wealth vanish. Contrary to the Schelling model, segregation is explicitly favored in the microscopic interactions. This Hamiltonian is furthermore equivalent to the field Hamiltonian $H[\phi]=\int \nabla \phi(\mathbf{x})^2 \dd \mathbf{x}$ in continuous space, whose equation of motion is the steady state diffusion equation. 

\begin{figure}[htbp]
\begin{center}
\centering
\includegraphics[scale=0.37]{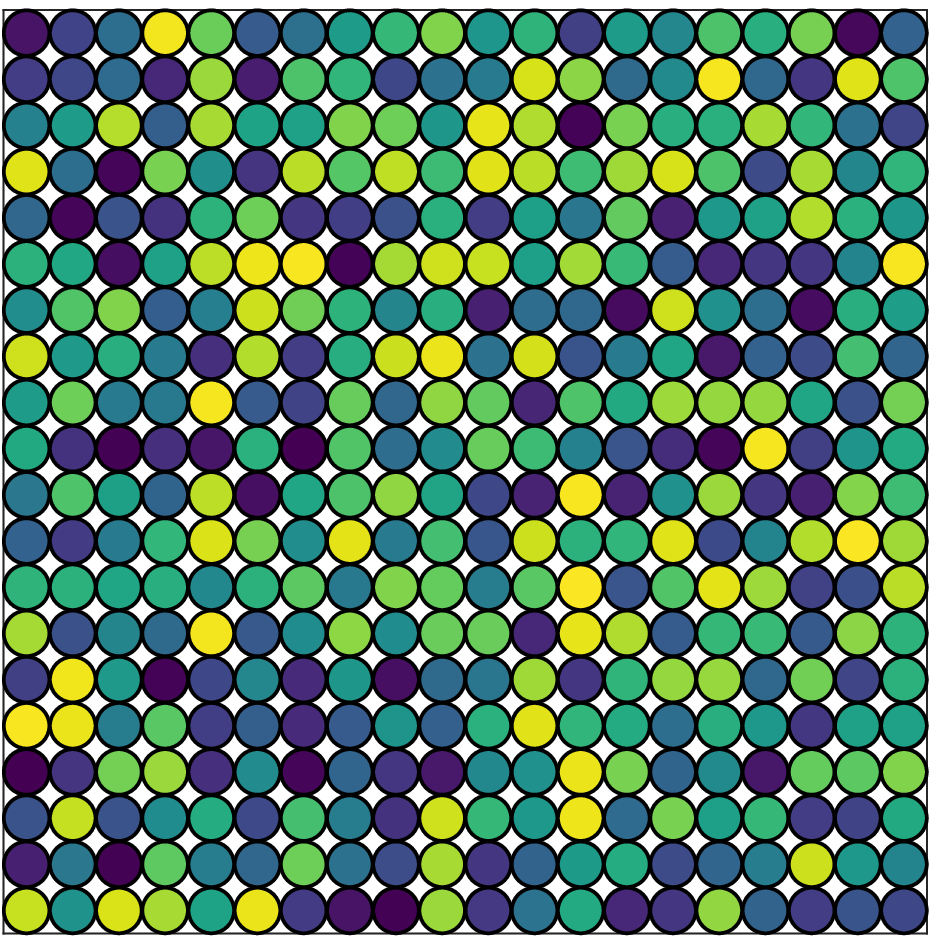} \quad
\includegraphics[scale=0.37]{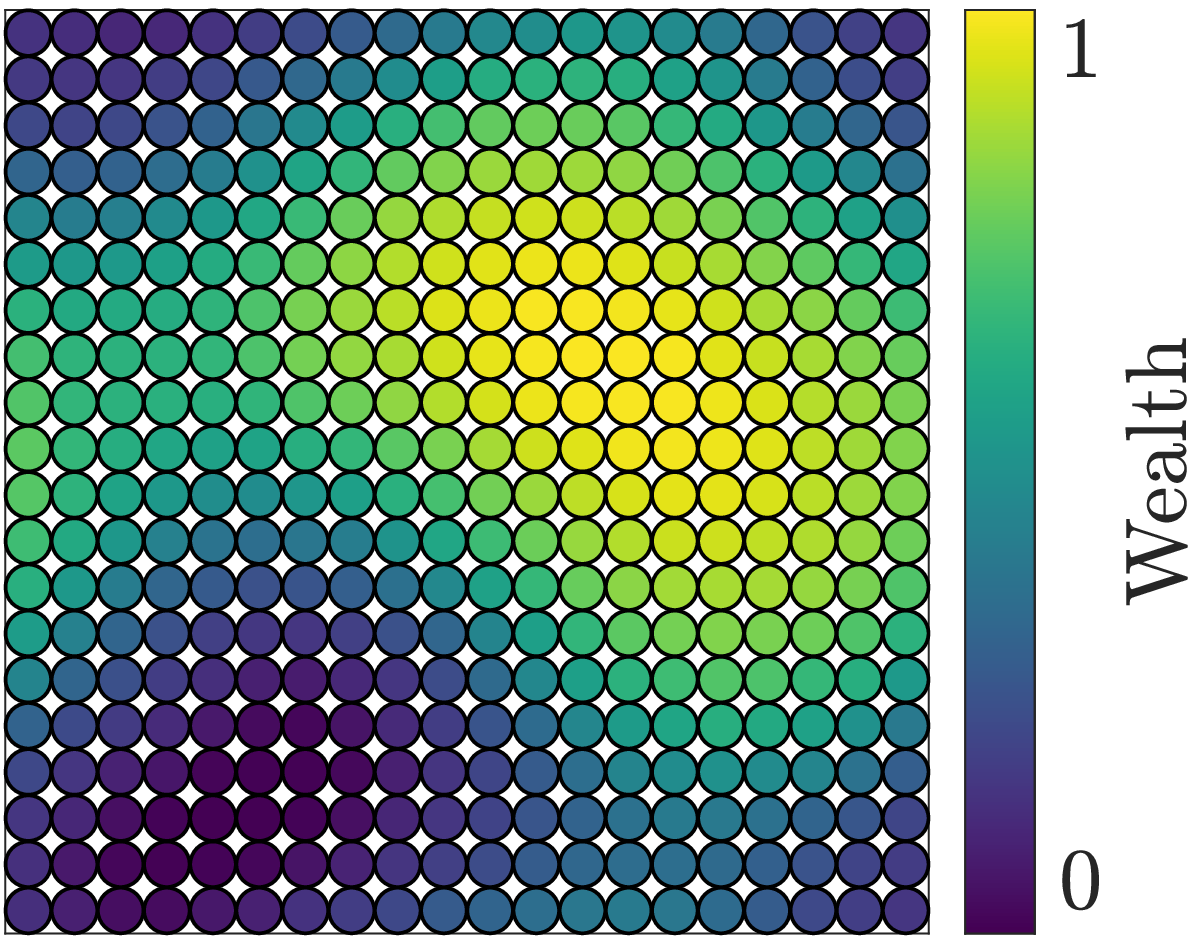}
\caption{Initial lattice (left) and typical steady state (right) reached by agent movements determined by Eq. \eqref{eq:metropolis} at $T=0$. For visual clarity, agents' individual wealth is fixed in this illustration; they move but do not exchange.}
\label{fig:lattice_illustration}
\end{center}
\end{figure}

A simulation starts with an $N\times N$ lattice filled with agents of wealth $m_i^{(t=0)}=1$. Unlike the Schelling model, agents do not move into vacant sites, the number of which would introduce an additional parameter \cite{gauvin_2009}. Instead, we propose at each time step a transition from state $\alpha$ to $\beta$, attained by switching two random agents in the lattice \footnote{Agents thus move over an arbitrary distance without any cost of moving. Another model, in which agents only switch places with nearest neighbours, did not result in large scale segregation at $T=0$.}. The transition is accepted with a probability given by the Metropolis-Hastings rule,
\begin{equation}
\mathcal{P}(\alpha\to \beta)=\textnormal{min} \left[1,\exp{\frac{-\Delta E}{T}}\right],
\label{eq:metropolis}
\end{equation}
where $\Delta E = E_{\beta} - E_{\alpha}$. These transitions are equivalent to Kawasaki kinetics for magnetic systems, which sample the equilibrium distribution at temperature $T$ while conserving the total magnetization \cite{kawasaki_1966}. When $T>0$, transitions increasing the energy of the system are realized with finite probability. As in earlier work \cite{stauffer_solomon_2007,sahasranaman_2016}, these fluctuations are interpreted here as external forces, causing agents to perform a move which violates their homophilic preference. This includes coincidences like personal quarrels or natural disasters, but also the consequences of policy promoting the interaction between economic classes. Examples of this are improved access to education, efforts to activate the unemployed, or housing subsidies to achieve residential mixing \cite{stiglitz_2013,stauffer_solomon_2007}. Note that a move lowering the energy evaluated by the Hamiltonian in Eq. \eqref{eq:hamiltonian} lowers tensions in the whole society, thus increasing the \emph{common good}. As such, Eq. \eqref{eq:hamiltonian} cannot be interpreted as an individual cost or utility function in the economic sense. This is in contrast to the egoistic moves of the Schelling model, which increase the happiness of one agent but may lead to an overall decrease in happiness. Figure \ref{fig:lattice_illustration} shows the effect of these microscopic moves in a lattice of agents that move but do not exchange. In the long-time limit, any initially mixed society evolves to a state of segregation, exhibiting a macroscopic wealth gradient throughout the lattice.

To complete the model, we include kinetic exchanges. After moving, one of the two involved agents undergoes an exchange with one of its four nearest neighbours, according to the rules outlined in Sec. \ref{sec:kinetic_models}. This fully coupled approach of segregation dynamics and kinetic exchanges is, at the present time, a new addition to the field.

\section{Results}
\label{sec:results}
\begin{figure}[htbp]
\centering
\includegraphics[scale=1]{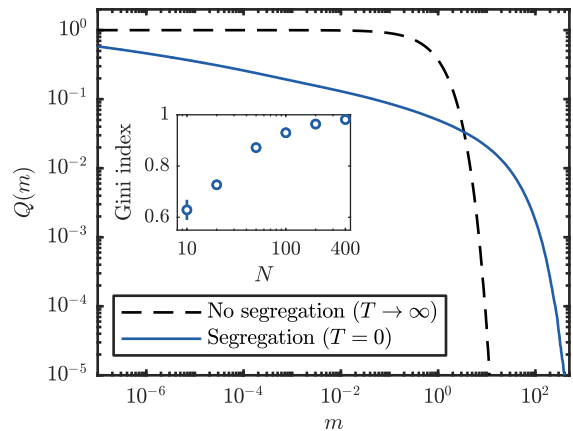}
\caption{Steady state distribution of the model outlined in Sec. \ref{sec:segregation}, using kinetic exchanges without saving propensity or taxation. The main figure is the result of 10 independent simulations in a $400\times 400$ lattice. The steady state at $T\to \infty$ was obtained by performing $t=10,000$ sweeps of $N^2$ microscopic moves, such that the distribution at time $t$ no longer deviates significantly from the one reached at time $t/2$. The $T=0$ result was obtained through a process of annealing detailed in Fig. \ref{fig:seg_temp_gini}. The inset shows the dependence of inequality on the lattice size $N$.}
\label{fig:seg_nosave_cdf}
\end{figure}

The solid line in Fig. \ref{fig:seg_nosave_cdf} shows the steady state distribution of the additive exchange model discussed in Sec. \ref{sec:kinetic_models}, excluding saving propensity or taxation and combined with fully deterministic segregation dynamics (i.e. $T=0$). For comparison, the dashed line shows the distribution at $T\to \infty$; this limit corresponds to a model without any restriction on interaction partners and follows a Boltzmann law as expected \cite{yakovenko_2009}. The global preference of surrounding oneself with neighbours of similar wealth status leads to a macroscopically segregated society and results in a distribution of wealth with a heavier tail (which is however no power law). Due to the spatial separation of poor and wealthy agents, the Gini index of inequality has increased from $0.5$ to approximately $0.98$ in the $400 \times 400$ lattice. Whereas poor agents in the unsegregated model could at any time interact with a rich agent and gain wealth quickly, they are now isolated from the vast portion of wealth in the economy. As such, the segregation mechanism impedes economic mobility by restricting access of poor agents to wealthy interaction partners.

\subsection{Saving propensity}
\begin{figure}[htbp]
\centering
\includegraphics[scale=1]{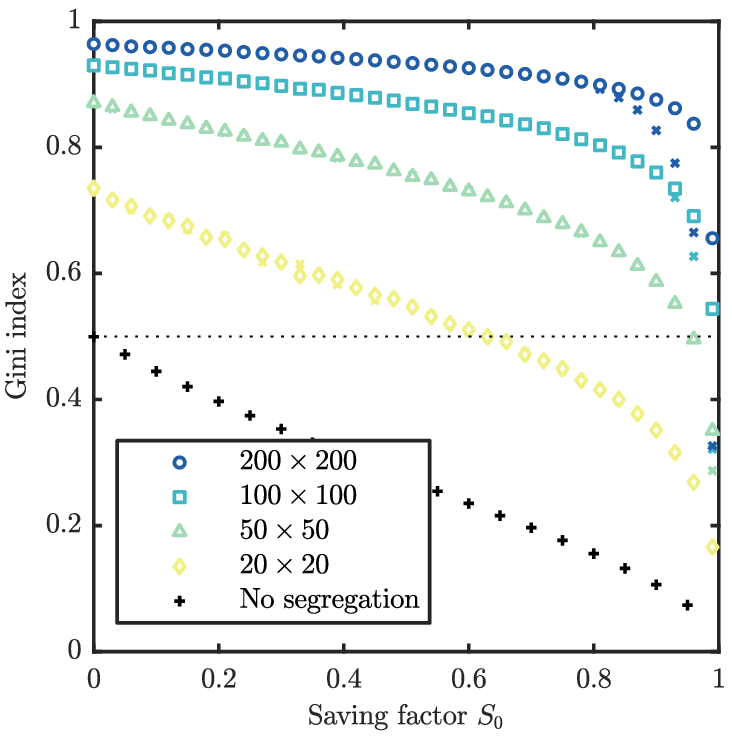}
\caption{Gini index of inequality in the model with uniform saving factor $S_0$ and segregation dynamics at $T=0$. Data points (o, $\scriptstyle\square$, $\scriptstyle\triangle$, $\diamond$, $+$) represent the distribution reached after 100,000 sweeps in lattices of the given size. The corresponding distributions after 10,000 sweeps (x) indicate the slow convergence in the limit of high saving propensity. Error bars representing the standard deviation of 10 independent simulations are smaller than the markers and were omitted.}
\label{fig:seg_S0_gini}
\end{figure}

\begin{figure}[htbp]
\centering
\includegraphics[scale=1]{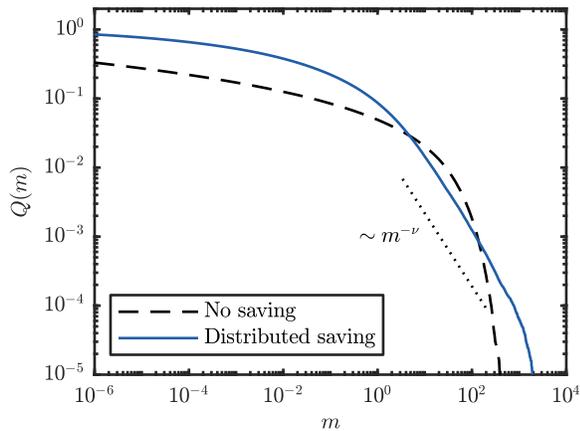}
\caption{Steady state distribution of the model with distributed savings and segregation dynamics at $T=0$. The figure is the result of 10 independent simulations in a $400\times 400$ lattice. The dashed black line is identical to the solid blue line in Fig. \ref{fig:seg_nosave_cdf}. The dotted line is a power law of exponent $\nu=1$.}
\label{fig:seg_sdist_cdf}
\end{figure}

Figure \ref{fig:seg_S0_gini} shows the Gini index of inequality for different values of the saving factor $S_0$ in the model with uniform savings (Sec. \ref{sec:saving_model}). A linear decrease to perfect equality can be noted in the absence of segregation, consistent with results reported in \cite{chakraborti_2000}. On a 2D lattice at $T=0$, this behaviour only remains for small societies. When the lattice size and hence the spatial concentration of wealth is increased, individual saving behaviour becomes a less effective strategy for combatting inequality. In a $200\times 200$ lattice, retaining $90\%$ of one's wealth in each transaction only results in a decrease of inequality from $G=0.96$ to $G=0.88$. Thus, in the presence of wealth-based segregation, saving behaviour alone doesn't suffice to produce levels of inequality consistent with empirical distributions such as Fig. \ref{fig:belgian_income}. Additionally, we find convergence towards the steady state is slowed down significantly in a large lattice and in the limit of high saving propensity.
The steady state distribution of the model with distributed savings and segregation is shown in Fig. \ref{fig:seg_sdist_cdf}, in comparison to the model without savings. The top percentiles obey a power law with exponent $\nu = 1.058 \pm 0.037$, consistent with the value $\nu=1$ reported in \cite{chatterjee_2004} for the same exchange model without segregation, but not in agreement with the empirical distribution in Fig. \ref{fig:belgian_income}. Patriarca et al. \cite{patriarca_2005} explained this power law behaviour by the occurrence of hoarders with a high saving factor who accumulate wealth. In particular, they found a relation between agents' wealth and saving factor of the form $\expval{m(S)}\sim \left( 1-\expval{S}\right)^{-1}$. We find a similar positive correlation, which is however exponential for the majority of poor agents, and follows the above form only in the top percentiles.

\subsection{Redistributive taxation}
\begin{figure}[htbp]
\centering
\includegraphics[scale=1]{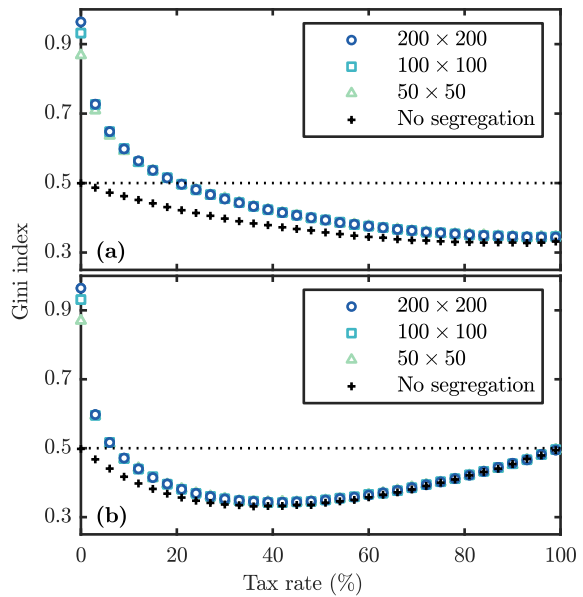}
\caption{Gini index of inequality in the model with redistributive income (a) and wealth (b) taxation, including segregation dynamics at $T=0$. Data points represent the steady state distribution reached after 2000 sweeps in lattices of the given size. Error bars representing the standard deviation of 10 independent simulations are smaller than the markers and were omitted.}
\label{fig:seg_tax_gini}
\end{figure}

The addition of a redistributive tax (Sec. \ref{sec:tax_model}) has a stronger effect on inequality. Both for income and wealth taxes, Fig. \ref{fig:seg_tax_gini} shows a sharp initial decrease of inequality with increasing tax rate. Whereas inequality in absence of taxation is highly dependent on the spatial concentration of wealth and thus on the lattice size, this dependence disappears asymptotically in the limit of high taxation: global redistribution transcends the spatial segregation.
Inequality decreases monotonically with increasing income tax, shown in Fig. \ref{fig:seg_tax_gini}(a),  converging to a finite value $G\approx 0.34$ when $100\%$ of each exchange is redistributed. While no one in this extreme scenario can gain wealth except through redistribution, each interaction still implies a random loss for one of the involved agents, preventing a state of perfect equality. We find that an income tax of 24\% reproduces the inequality of the empirical distribution in Fig. \ref{fig:belgian_income}. This is in contrast to an unsegregated economy, where $G=0.5$ is reached by additive exchanges without redistribution. Also in the case of wealth taxation, Fig. \ref{fig:seg_tax_gini}(b), inequality remains finite when all wealth is redistributed, since this scenario is equivalent to randomly reducing one agent's wealth to zero at each time step. Interestingly, this limit yields exactly the natural value $G\approx 0.5$ for the case of additive exchanges in an unsegregated economy. Contrary to income taxation, inequality is now minimized for a finite wealth tax rate of approximately $40\%$. The necessity of taxation to achieve levels of inequality found in empirical distributions, which are likewise shaped by income taxes, provides an indication for the role of wealth-based segregation in economic inequality.

\subsection{Thermal fluctuations}
\begin{figure}[htbp]
\centering
\includegraphics[scale=1]{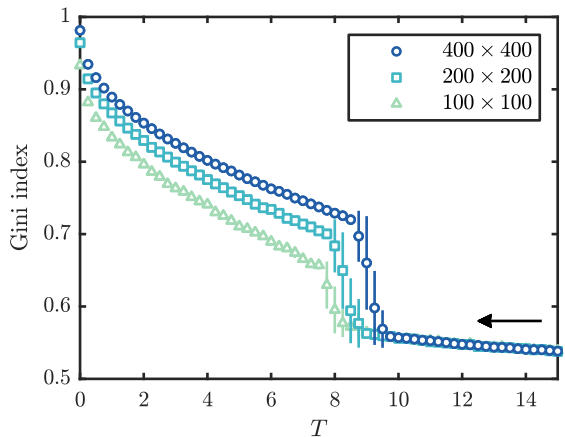}
\caption{Gini index of inequality in the model without saving or taxation, and segregation dynamics at finite temperature. Data points represent the steady state distributions reached during a process of annealing from high to low $T$, in which the system relaxes to its steady state after 1000 sweeps at a given temperature, before being cooled down by $\Delta T=0.25$. Error bars represent the standard deviation after 10 independent simulations. The arrow indicates the direction in which the phase boundary is crossed.}
\label{fig:seg_temp_gini}
\end{figure}

So far we have concentrated on fully deterministic segregation, that is, the $T=0$ limit of the transition rule given by Eq. \eqref{eq:metropolis}. This means an agent will never move to a location perceived as disadvantageous. We now turn our attention to thermal fluctuations in these homophilic moves. As seen from Eq. \eqref{eq:metropolis}, the effect of a non-zero temperature is to allow for such disadvantageous transitions.
Figure \ref{fig:seg_temp_gini} shows the Gini index as a function of $T$ for the simplest model without saving or taxation. The overall decrease of inequality with rising temperature is to be expected, since noise is added to a mechanism which increases inequality. At a critical temperature $T_C$ we observe a sudden decrease in inequality, which becomes sharper and with bigger fluctuations as the thermodynamic limit is approached, resembling a second order phase transition. The exponential shape and resulting Gini index of the empirical distribution shown in Fig. \ref{fig:belgian_income} indicate that, in the presented framework, the Belgian economy corresponds to the unsegregated phase.

\begin{figure}[htbp]
\centering
\includegraphics[scale=1]{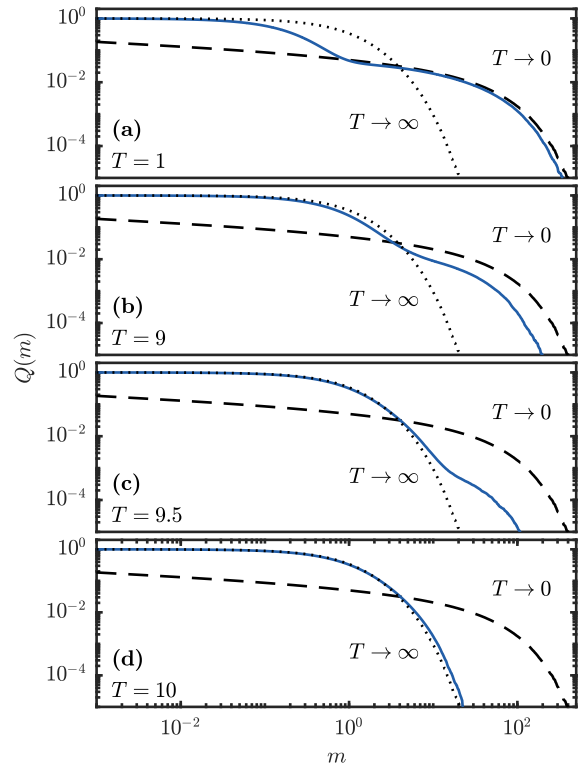}
\caption{Steady state distribution of the model with segregation dynamics at finite temperature. For comparison we also show the distributions at $T=0$ and $T\to \infty$, equivalent to those in Fig. \ref{fig:seg_nosave_cdf}. Each distribution corresponds to an equilibrium point of the $400 \times 400$ curve in Fig. \ref{fig:seg_temp_gini}.}
\label{fig:seg_temp_cdf}
\end{figure}

More insight into the nature of this transition is obtained from the distribution of wealth around the critical temperature. Figure \ref{fig:seg_temp_cdf} shows the cumulative distribution function for $T\ll T_C$, $T< T_C$, $T=T_C$ and $T> T_C$. From Figure \ref{fig:seg_temp_cdf}(a) we conclude that at $T\ll T_C$, the distribution can roughly be divided into two regimes. Among poor agents, differences in wealth are relatively small. As such, the quantity $\Delta E$ is small for moves involving poor agents. A low but finite temperature in this case suffices to let the exponential in Eq. \eqref{eq:metropolis} approach unity. For moves involving rich agents, $\Delta E$ can take on larger values and hence a higher temperature is required for thermal fluctuations to have a significant effect on the resulting distribution. This explains why the poor majority desegregates at low temperature and follows the distribution for $T\to \infty$, while the top percentiles still follow the $T=0$ distribution. As can be seen from Figure \ref{fig:seg_temp_cdf}(b)-(d) the phase transition to lower inequality is related to a sudden desegregation among the wealthy elite. The resulting mixing of classes increases interactions among them, which is seen to reduce inequality in the global distribution of wealth.

\section{Conclusion and outlook}
\label{sec:conclusion}
We have shown that when economic agents, exchanging wealth through kinetic collisions on a lattice, continuously switch places in an attempt to minimize local differences in wealth, inequality is increased dramatically due to the macroscopic separation of rich and poor agents. Inequality in itself isn't necessarily perceived as problematic, so long as there is economic mobility allowing for rags-to-riches stories. This has been referred to as the \emph{prospect of upward mobility} (POUM) hypothesis \cite{benabou_ok_2001}. Our toy model shows how both phenomena can be intricately linked, with reduced economic mobility leading to a more unequal society on the whole.

Strategies of combatting inequality on a microscopic level, such as individual saving behaviour, prove ineffective in the segregated economy due to limited exposure of poor agents to wealthy interaction partners. On the contrary, a global redistributive income or wealth tax transcends the spatial inhomogeneity and leads to a strong decrease in inequality, quickly nullifying the effect of segregation. Thermal fluctuations in the moving dynamics also decrease inequality, with a sharp phase transition at a critical temperature caused by a sudden desegregation among the wealthy elite. The possibility of a sudden decrease of inequality, caused by a minor change in external factors, suggests that small measures of economic policy may suffice to reduce inequality, if only they manage to carry the society across such a transition.

In our approach, kinetic exchanges were restricted to nearest neighbours on a 2D lattice, inspired by the abundant literature on residential segregation. A more realistic model of economic interactions should take into account that not all agents support an equal number of links. Hence, the extension to more general network structures is a natural next step. By displacing links according to the proposed transition rule, wealth-based segregation could be used to generate the structure of a socio-economic network. The rules obeyed in kinetic exchanges have been kept minimal as well. In the past, agent-based models have been developed including stochastic growth, debt, firms, banks, tradeable commodities, progressive taxation and other refinements \cite{chatterjee_2007, yakovenko_2009, bouchaud_2000, de_oliveira_2017}. Our findings suggest a simple way of incorporating social preferences and economic mobility into agent-based studies of wealth and inequality.

\begin{acknowledgments}
L. F. gratefully acknowledges support in the form of a Ph.D. fellowship from the Research Foundation - Flanders (FWO), project 11E8120N.
\end{acknowledgments}

\section*{Author contribution statement}
J.T. initiated and supervised the project. L.F. devised the computational model, performed simulations and wrote the initial draft of this paper. Both authors contributed equally in the interpretation of results and revision of the manuscript.

\end{document}